\documentclass[11pt]{article}
\usepackage{graphicx}
\usepackage{epstopdf, epsfig}
\usepackage{amssymb,amsmath}
\usepackage{color}
\usepackage{comment}
\usepackage{caption}
\usepackage{subcaption}
\usepackage{comment}
\usepackage{rotating}
\usepackage{natbib}

\newtheorem{lemma}{Lemma}

\newcommand{\ka}{\ensuremath{\bold{k}}}
\newcommand{\x}{\ensuremath{\bold{x}}}
\newcommand{\ele}{\ensuremath{\bold{l}}}
\newcommand{\M}{\ensuremath{\bold{M}}}
\newcommand{\ep}{\ensuremath{\varepsilon}}
\DeclareMathOperator{\sech}{sech}

\title{A three dimensional Dirichlet-to-Neumann operator for water waves over topography}

\author{D. Andrade and A. Nachbin}
\date{}

\begin{document}

\maketitle

\begin{abstract}
Surface water waves are considered propagating over highly variable non-smooth topographies. For this  three dimensional 
problem a Dirichlet-to-Neumann (DtN)
operator is constructed  reducing the numerical modeling and evolution to the two dimensional free surface.  The
corresponding Fourier-type operator is defined through a matrix decomposition. The topographic component of the 
decomposition requires special care and a Galerkin method is provided accordingly. One dimensional numerical 
simulations, along the free surface, validate the DtN formulation in the presence of a large amplitude, 
rapidly varying topography.  An alternative, conformal mapping based, method is  used for benchmarking. A two dimensional
simulation in the presence of a Luneburg lens (a particular submerged mound) illustrates the accurate performance of the  
three dimensional DtN operator.
\end{abstract}

\section{Introduction}

Water wave dynamics in coastal regions, where surface waves are affected by depth variations, is a problem of great 
interest from the environmental point of view as well as due to its mathematical (formulation, theory and computational)
diversity. It is a long  lasting source of problems, with still many challenges. 
The problems are very complex, in particular in the presence of a bottom 
topography. Some of these problems are related to the Euler (potential theory) equations, while others to 
its reduced models, such as the weakly nonlinear, weakly dispersive Bousssinesq system. 
The topography may come into the problem 
in a geometric fashion
as a variable boundary or, as a variable coefficient (in an algebraic form).  
Inclusion of  topographic features, in either way, is a relevant physical ingredient of current research interest and has  been
addressed  from  the Analysis viewpoint, as well as regarding 
computational modeling and related issues. 

Non-local water wave formulations make use of the harmonic properties of the velocity potential. In incompressible and 
irrotational flows this has been know for quite some time, in particular in two dimensions. Boundary integral equations
and complex function theory, by the Cauchy  integral formula, have been widely used. The literature is abundant,
accessible and therefore
we will refrain from citing just a few authors. 
Nevertheless the papers cited herein are a  good sources of references. 
More recently non-local formulations through the Dirichlet-to-Neumann operator
became popular, both theoretically and numerically. The operator collapses all information onto the free surface.
The dimension of the problem is reduced. For example, the dynamics in a three dimensional fluid body is reduced
to computing  only with information along the two dimensional free surface.  

The non-local operators here considered  arise in the form of Fourier integrals: 
\begin{equation}
F[f](x) = \int\!\!\! \int e^{ik(x-x^\prime)} w(\ka) f(x^\prime) dx^\prime dk.
\end{equation}  
The Fourier operator $F$  acts on $f$, where $w(\ka)$ is the Fourier multiplier (symbol).  If the multiplier
is a polynomial in $k$, then $F$ is a differential operator. 
Due to the presence of the topography in the present formulation, the symbol will be 
$x$-dependent. 

An early derivation is due to \citet{Paul1998}, where the author proposes a Fourier integral equation for the water wave problem. 
Milewski uses the two (linearly independent) hyperbolic functions for harmonically extending 
boundary data into the fluid domain.
By expanding the symbol  appropriately the Fourier integral, to leading order, characterizes a differential 
operator. Milewski then recovers the standard long-wave (asymptotic) equations such as the Benny-Luke, the 
Boussinesq, and the KdV equations. His models incorporate  topographical effects through 
variable coefficients and forcing terms. These approximations are valid when the topographical 
variations are slowly varying  with respect to the wave field.

Recently the non-local formulation proposed by \cite{AFM}  has called attention as an  
alternative water wave modeling. The authors 
derive a set of global relations that couple free surface equations together with Fourier components. The Fourier 
components expressed through hyperbolic functions, when expanded, given rise to differential operators defined
as powers of the wavenumber $k$ in Fourier space. Hereafter the formulation proposed in 
\cite{AFM} will be called the AFM formulation.
\citet{OliverasDeconinck} used the AFM formulation to compute periodic traveling wave solutions. 
Their work  represents one of the first steps towards the use of the AFM (non-local) equations
in the  numerical simulation of water waves.
They studied wave stability, showing that it is possible to solve the non-local equations accurately. 

The AFM formulation has been considered in other water wave problems, as for example by
\citet{AblowitzHaut} for internal waves and by \cite{AshtonFokas} for waves with vorticity.

\citet{VasanDeconinck} made a contribution in the context of time dependent AFM simulations, 
in the presence of a topography. Their goal was the recover the bottom profile 
from surface measurements. This task required  an accurate time dependent numerical modeling.
Long, small amplitude waves were used for the bottom reconstruction. 
The dispersion parameter $\mu$ (the ratio between wavelength $\lambda$ and depth $h$) was set to $0.016$ 
in most of the simulations.
Attention was paid on the number of Fourier modes needed to reconstruct the bottom accurately. 
The smooth obstacles considered were wavy bottoms and Gaussian mounds, among others. Their amplitude 
ranged up to $25\%$ of the total depth of the fluid. 
In most cases the typical number of Fourier modes is about 12 for the bottom and around 60 for the wave. 
As  discussed in detail in \citet{VasanDeconinck}, the use of a large number of Fourier modes can become an issue, in 
particular in the presence of a topography and of moderate values of $\mu$.

A very detailed comparison between five different  
Dirichlet-to-Neumann formulations  is provided by \citet{WilkeningVasan}. 
They compare expansion methods against methods for solving the non-local equations directly. 
They focus on the two dimensional non-linear regime with a flat bottom.
Among others they consider the AFM-method as well as their dual formulation,
called the AFM$^*$-method. The formulation found in  \citet{Paul1998,CSNG} is namely an AFM$^*$-method. 
\citet{WilkeningVasan} propose the use of  arbitrary precision to deal with the ill-conditioned nature of the problem. 
Our formulation follows the line of \cite{Paul1998} (AFM$^*$ formulation) together with a Galerkin method,
leading to accurate results in the presence of highly variable topographies, including in three dimensions.

Non-local equations in the presence of topography also appear in the work of \citet{CSNG}. The authors define the 
respective Fourier multiplier operators as will be briefly described.

Having in mind their work  
from \citet{CraigSulem}, they  show that the topographical contribution is given by an extra term in the
Fourier representation. 
For simplicity, consider the Fourier multiplier operator-notation in two dimensions, ignoring for a
moment the time-dependence of the potential:
\begin{equation}
\phi(x,z) = \int\!\!\! \int e^{ik(x-x^\prime)}\frac{\cosh(k(z+h))}{\cosh(kh)} q(x^\prime) dx^\prime dk\equiv
\frac{\cosh((z+h)D)}{\cosh(h D)} q(x),
\end{equation}  
where $D\equiv i\partial x$.  It is easy to see, by separation of variables,  that the Fourier integral 
representation of the potential  satisfies Laplace's equation in a two dimensional strip, of height  $h$, together
with the following boundary conditions: a 
homogeneous Neumann condition at the bottom $z=-h$ and a Dirichlet condition $\phi(x,0)=q(x)$ at the top
boundary $z=0$. 
In three dimensions the notation becomes $D= (D_1,D_2)^T\equiv i(\partial x,\partial y)^T$. In the presence of a 
bottom topography $z=-(h+H(\x))$ the Fourier operator has an additional term in order to account for  the presence
of the topography $H(\x), ~\x=(x,y)$. \citet{CSNG} present their ansatz  by expressing the velocity potential as
\begin{equation}
\phi(\x,z)=\frac{\cosh((z+h)|D|)}{\cosh(h|D|)} q(\x) + \sinh(z|D|)L(H) q(\x).
\label{CSNG}
\end{equation}
The Fourier operator  $L(H)$ is defined implicitly  (see equation (2.23) and Appendix in \cite{CSNG}),
 through  a relation which contains two, topography dependent, Fourier operators $A(H)$ and $B(H)$:
\begin{eqnarray} 
A(H) =  \sinh(H(\x)|D|)\sech(h|D|), ~ ~ ~C(H)= \cosh((H(\x)-h)|D|),\\
 \mbox{with}~ B(H)=C(H)^{-1} \nonumber.
\end{eqnarray}

In their appendix, assuming that the bottom topography is small with respect to the reference depth,  
\cite{CSNG} give a series expansion for the topographical operator. 
The authors comment, in the conclusions,  that {\it ``Preliminary numerical results appear in
\cite{GuyenneNicholls}. More in-depth numerical computations constitute a substantial contribution in their own right(...)."} 
\citet{GuyenneNicholls} considered  expansions for the wave and the bottom topography operators. 
Their model was used to simulate the shoaling of solitary waves propagating over inclined planes. High-frequency 
instabilities were reported and treated with a low-pass filter for the velocity potential and wave elevation functions.
However it is not clear what is exactly the source of the instability. 
As reported by the authors,  it could be due to the ill-possedness of the truncation process that arises 
in the wave expansion, as recently investigated by \citet{Ambrose}, or  it could be due to the topographic operator
component?

In the present work we focus on the topographic component of the Fourier operator
which  is computed numerically as a matrix decomposition,
along the lines mentioned above.  Our ansatz for the topographic component is different from that 
displayed in (\ref{CSNG}) and follows the simpler expression given by \cite{Paul1998}.  
We take into account highly variable non-smooth topographies, 
which inhibits the use of a series expansion. A physically-motivated Galerkin method
acts only on low frequencies of the topographic component of the Dirichlet-to-Neumann
operator, keeping only modes that can indeed interact with the bottom variations. 
The velocity potential $\phi$ and the wave elevation $\eta$ are not filtered. 
Hence with the ingredients we considered the topographic-formulation, along the lines of \cite{Paul1998} and  \cite{CSNG}, 
performed very well for a range of nontrivial problems. These include large amplitude rapidly varying topographies, as well as
three dimensional landscapes, where the non-local formulation had not been tested before. 
Three dimensional simulations can be very time consuming and the Galerkin method played an important role. 

A recent work by \citet{Cathala} regards the analysis and two dimensional computations with the formulation provided by \cite{CSNG}. 
Cathala takes a second order expansion in both the wave  and the topographic operators. 

\citet{Cathala}  considers, in particular, the case of polygonal topographies: a submerged step that rises 
up to half the depth of the channel. He compares his result with that of \citet{Nachbin2003} and 
obtains a satisfactory comparison.
Based on conformal mapping, Nachbin considers a Boussinesq system in the presence of a polygonal topography.

In another recent work,  \citet{VargasPanayotaros} use the series expansion of \citet{CSNG} and deduce a 
Whitham-Boussinesq model with topography. The numerical modeling is two dimensional.
Their approach simplifies the contribution of the topographic operators 
by introducing a simpler non-local equation. They show that their operator captures the right asymptotic behaviour 
for the high wavenumbers. Their numerical scheme also uses a Galerkin approximation similar to ours and is tested for
a smooth submerged Gaussian mound.

The references mentioned  base their numerical methods on  series expansions of the Dirichlet-to-Neumann operator. 
However as pointed out by \citet{CSNG}, when the topography profile is $\mathcal{O}(1)$ a Taylor expansion 
is not what one should seek. 

In summary, the technique here presented is validated 
for solving the nontrivial linear water wave problems in three dimensions. The method uses a non-local 
formulation that reduces the evolution system to a two dimensional surface. As our main 
goal, a Dirichlet-to-Neumann (DtN) operator
is constructed in the presence of large amplitude, non-smooth 
topographies retaining, along the two dimensional surface, information from the vertical flow structure.  
A Galerkin method in Fourier space is used to make the topographic component of the DtN operator
amenable to accurate computations. 

This article is organized as follows. In section 2 we  introduce the potential theory equations and the associated
elliptic boundary value problem for constructing the Dirichlet-to-Neumann operator over a flat bottom. 
Section 3 considers  the variable-depth non-local formulation and the associated decomposition for the 
three dimensional DtN operator.
Section 4 presents several water wave simulations in two and three dimensions. 
The conclusions are given in section 5.

\section{Governing equations}

\subsection{Water wave equations}
Potential theory is the classical hydrodynamic model that describes the motion of an incompressible, irrotational flow under the action of gravity (\cite{Whitham}):
\begin{eqnarray}
	\phi_{xx}+\phi_{yy}+\phi_{zz}&=& 0,\ \ \text{in $-b(\x)<z<\eta(\x,t)$.}\label{S1:Eq01}\\
	\phi_{z} - \phi_{x}b_x - \phi_{y}b_y &=& 0,\ \ \text{on $z=-b(\x)$.}\label{S1:Eq02}\\
	\eta_t + \phi_x\eta_x + \phi_y\eta_y - \phi_z&=& 0 ,\ \ \text{on $z=\eta(\x,t)$.}\label{S1:Eq03}\\
	\phi_t + g\eta +\tfrac{1}{2}\left(\phi_x^2+\phi_y^2+\phi_z^2\right) &=& 0,\ \ \text{on $z=\eta(\x,t)$.}\label{S1:Eq04}
\end{eqnarray}
In these equations $\x = (x,y)$ denotes the horizontal variables, $g$ is the acceleration due to gravity, $\eta$ denotes the height of the surface wave, $\phi$ is the velocity potential of the fluid flow and finally $b$ denotes the 
impermeable bottom topography. We assume that $\eta$, $\phi$ and its derivatives vanish at infinity.

The bottom of the fluid domain is described as
\begin{equation}\label{S1:Eq05}
b(\x) = h + H(\x),
\end{equation}
where $h$ is a positive constant and the depth-variation function $H$ is bounded: 
\begin{equation}\label{S1:Eq06}
|H(\x)|<c<h, ~~c > 0.
\end{equation} 
This restriction implies that no islands nor beaches are present in the fluid domain. But it allows for high amplitude, non-smooth obstacles, as will be presented herein.

\subsubsection{Non-dimensional form}

It is customary to work with a non-dimensional form of the equations. This can be accomplished by replacing all the variables in equations (\ref{S1:Eq01})-(\ref{S1:Eq04}) with primes  and  making the following substitutions:

\begin{equation}\label{S1:Eq07}
\begin{aligned}
x' = lx, \ \ \ & y' = ly, \ \ \ z' = hz \ \ \ t' = \frac{l}{\sqrt{gh}}t,\\
\eta' = a\eta,\ \ \ & \phi' = a\sqrt{gh}\frac{l}{h}\phi, \ \ \ H' = hH.\\
\end{aligned}
\end{equation}
In these new variables $l$ denotes a typical length-scale for both horizontal directions, $h$ denotes a typical depth of the fluid domain, $a$ denotes a typical amplitude of the wave and $\sqrt{gh}$ is the typical speed of long waves propagating over an ocean of constant depth $h$. The following dimensionless parameters arise: $\mu = {h}/{l}$ and $\ep = {a}/{h}$.

The dimensionless equations are 
\begin{eqnarray}
\mu^2\left(\phi_{xx}+\phi_{yy}\right)+\phi_{zz} &=& 0,\ \ \ \ \text{in $-(1+H(\x))<z<\ep\eta(\x,t)$,}\label{S2:Eq08}\\
\phi_z -\mu^2(\phi_{x}H_x+\phi_{y}H_y)&=&0,\ \ \ \ \text{on $z=-(1+H(\x))$,}\label{S2:Eq09}\\
\eta_{t} + \ep(\phi_x\eta_x + \phi_y\eta_y) - \tfrac{1}{\mu^2}\phi_z&=&0,\ \ \ \ \text{on $z = \ep\eta(\x,t)$,}\label{S2:Eq10}\\
\phi_t + \eta +\tfrac{\ep}{2}\left(\phi_x^2+\phi_y^2+\phi_z^2\right) &=&0,\ \ \ \ \text{on $z = \ep\eta(\x,t)$.}\label{S2:Eq11}
\end{eqnarray}

\subsection{Linear model}
The potential theory equations have two main sources of difficulties, namely, the effects of the underwater topography and the nonlinear effects. Our objective is to study the effects due to the bottom topography.  Therefore
we keep the nonlinear effects aside and consider 
\begin{eqnarray}
	\mu^2(\phi_{xx}+\phi_{yy})+\phi_{zz}&=& 0,\ \ \text{in $-1-H(\x)<z<0$,}\label{S2:Eq12}\\
	\phi_{z} - \mu^2(\phi_{x}H_x + \phi_{y}H_y) &=& 0,\ \ \text{on $z=-1-H(\x)$,}\label{S2:Eq13}\\
	\eta_t  - \tfrac{1}{\mu^2}\phi_z &=& 0 ,\ \ \text{on $z=0$,}\label{S2:Eq14}\\
	\phi_t + \eta  &=& 0,\ \ \text{on $z=0$.}\label{S2:Eq15}
\end{eqnarray}
In terms of physical scales the linear dispersive system of equations is well suited for the investigation of 
small amplitude surface waves,  as for instance, a tsunami that propagates offshore. 
See \citet{ArcasSegur} for further details on this matter.

\subsubsection{Time evolution equations}
The three-dimensional linearized water wave equations can be put into an equivalent two-dimensional form,
restricted to the free surface,  by means of the Dirichlet-to-Neumann operator. We introduce the variable $q(x,y,t) = \phi(x,y,0,t)$ 
and denote by $G[q]$ the compatible Neumann data 
regarding the respective elliptic problem. The linearized water wave equations become:
\begin{eqnarray}
\eta_t &=& \tfrac{1}{\mu^2}G[q],\label{S1:Eq16}\\
q_t &=& -\eta.\label{S1:Eq17}
\end{eqnarray}

In the next section we define the corresponding elliptic problem and show how to compute the Dirichlet-to-Neumann operator $G$,  
which captures features of the vertical structure of the flow including the presence of  a highly
variable topography.

Once the operator $G$ is constructed for a given bottom topography $H$, we proceed to solving the evolution equations (\ref{S1:Eq16}) and (\ref{S1:Eq17}) with a fourth order Runge-Kutta method. We will obtain accurate results with this formulation for 
three dimensional flows  in the presence of 
a highly variable topography, a regime not yet explored with the DtN operator.
\subsection{The elliptic  boundary value problem for water waves}


In this section we define the elliptic boundary value problem associated with the linear equations for gravity waves over topography. We assume that the wave dynamics takes place inside a bounded region of space. 
For the free surface we choose a sufficiently large square of sides $L$, that contains the region of interest. Without loss of generality we impose periodic boundary conditions at the ends of this region.

Under these considerations, let 
\begin{equation}\label{S2:Eq01}
D = \{(x,y,z)\in\mathbb{R}^3 \mid 0 <x< L,\ 0 <y<L ,\ \text{and\ } -1-H(\x)<z<0\},
\end{equation}
be the fluid domain and consider the following elliptic boundary value problem: 
\begin{eqnarray}
\mu^2(\phi_{xx}+\phi_{yy})+\phi_{zz}&=& 0,\ \ \text{in $D$,}\label{S2:Eq02}\\
\phi(\x,z) &=& q(\x),\ \ \text{on $z=0$,}\label{S2:Eq03}\\
\phi_{z} - \mu^2(\phi_{x}H_x + \phi_{y}H_y) &=& 0,\ \ \text{on $z=-1-H(\x)$.}\label{S2:Eq04}
\end{eqnarray}
The solution to this problem determines the dimensionless velocity potential $\phi$ inside the fluid domain $D$. The problem is linear and the domain $D$ is fixed. Therefore the solution only depends on the boundary value $q$ at any given time $t$.
Hence time will be temporarily omitted in the presentation that follows.

Regarding the solvability of the above elliptic boundary value problem, \citet{Lannes2005} proved that in the presence of a smooth bottom and appropriate decaying conditions at infinity, the problem always admits a unique smooth solution $\phi(\x,z)$.

For our evolution equations a quantity of great interest is the vertical speed of the fluid ($\phi_z$) along 
the linearized free surface $z = 0$. This term has the information of an underlying harmonic function $\phi$, satisfying a Neumann 
condition on the variable bottom boundary. 
It  is obtained through the Dirichlet-to Neumann (DtN) operator applied to the function $q$, which mathematically 
reduces  one dimension of the problem.  The respective Fourier-type DtN operator 
collapses the three dimensional dynamics onto  free surface equations. We use the notation
\begin{equation}\label{S2:Eq05}
G[q] = \phi_z,\ \ \ \text{on $z = 0$}.
\end{equation}

\subsubsection{Flat bottom case}
In order to motivate our strategy for the constructing the variable-depth Dirichlet-to-Neumann operator, 
we first discuss the ($H=0$) flat bottom case. 
In the flat bottom case, Fourier analysis immediately gives us an explicit formula for the solution $\phi$ of equations (\ref{S2:Eq02})-(\ref{S2:Eq04}):
\begin{equation}\label{S2:Eq06}
\phi(\x,z) = \sum_{\ka\in\Lambda}e^{i\ka\cdot\x}\hat{q}(\ka)\frac{\cosh(\mu k(z+1))}{\cosh(\mu k)},
\end{equation}
where $\ka = (k_1,k_2)$, $k=\sqrt{k_1^2+k_2^2}$. 
The Fourier coefficient of the Dirichlet data $q(\x)$ is
defined by
\begin{equation}\label{S2:Eq07}
\hat{q}(\ka) = \int_{T^2}e^{-i\ka\cdot\x}q(\x)\ d\x,
\end{equation}
where we have denoted the undisturbed free surface ($z=0$) by 
\begin{equation}\label{S2:Eq08}
T^2 = \{\x\in\mathbb{R}^2 \mid 0 <x< L,\ \text{and\ } 0 <y<L \}.
\end{equation}
Doubly periodic functions in $T^2$ have their Fourier spectrum in the set
\begin{equation}\label{S2:Eq09}
\Lambda = \tfrac{2\pi}{L}\mathbb{Z}\times\tfrac{2\pi}{L}\mathbb{Z}.
\end{equation}
We denote the set of non-zero wavenumbers by 
\begin{equation}\label{S2:Eq10}
\Lambda^* = \{ (k_1,k_2)\in\Lambda \mid k_1^2 + k^2_2 > 0\}.
\end{equation}

From (\ref{S2:Eq06}) we  compute the vertical derivative of $\phi$ along the undisturbed free surface:
\begin{equation}\label{S2:Eq11}
G[q] := \phi_z(\x,0) = \sum_{\ka\in\Lambda^*}e^{i\ka\cdot\x}\hat{q}(\ka)\mu k\tanh(\mu k)\ d\ka.
\end{equation}
Formula (\ref{S2:Eq11}) defines the Dirichlet-to-Neumann operator for channels with a flat bottom. 

\section{Variable-depth non-local formulation}

In order to study the elliptic boundary value problem (\ref{S2:Eq02})-(\ref{S2:Eq04}) we revisit an alternative non-local formulation of the problem which captures the effects of the bottom topography through the vertical velocity of the fluid, at the linearized free surface, $(z=0)$. As mentioned in the introduction, this non-local formulation was proposed in the work of \citet{Paul1998}, and more recently by \citet{CSNG} and \citet{AFM}.
The work of \citet{VasanDeconinck} was one of the first articles in which the AFM formulation was used with 
bathymetry variations. The authors studied the possibility of reconstructing the sea-bed topography  
based on surface-wave measurements.

Our version of the AFM$^\ast$ formulation is linear and in three dimensions, as opposed to that of \citet{VasanDeconinck}  
which was nonlinear and in two dimensions. Our goal is to take one step forward by 
modeling complex (not necessarily smooth) topographies that involve large and rapid variations and, as mentioned, 
perform simulations with fully three dimensional configurations. 
This is the first time that the non-local method is used under these conditions.


The non-local variable-depth formulation arises from the following representation formula, for the solution $\phi(\bold{x},z)$ to the elliptic boundary value problem (\ref{S2:Eq02})-(\ref{S2:Eq04}):
\begin{equation}\label{S3:Eq01}
\phi(\bold{x},z) = \hat{q}(0) + \sum_{\ka\in\Lambda^*}e^{i\ka\cdot\x}\left[\hat{q}(\ka)\frac{\cosh(\mu k(z+1))}{\cosh(\mu k)} + X_\ka\frac{\sinh(\mu kz))}{k\cosh^2(\mu k)}\right].
\end{equation}
 In this representation formula (\ref{S3:Eq01}), the complex coefficients $X_\ka$ have to be determined from 
 the bottom boundary condition (\ref{S2:Eq04}). 
 Notice that if $X_\ka$ vanished for all $\ka$, then we would recover (\ref{S2:Eq06}), the solution of the flat bottom case. 
 Therefore $X_\ka$ contains  information related to the geometry of the topography. We will call it the topographic coefficients.

The impermeability boundary condition (\ref{S2:Eq04}) yields the following linear relation between the 
topographic coefficients and the (given) Dirichlet data $q$:

\begin{equation}\label{S3:Eq02}
\sum_{\ka\in\Lambda^*} \hat{q}(\ka)\nabla \cdot \left[ e^{i\ka\cdot\bold{x}} \frac{\sinh(\mu k H(\bold{x}))}{\cosh(\mu k)}\frac{\ka}{k}\right] = \sum_{\ka\in\Lambda^*} X_\ka\nabla\cdot \left[ e^{i\ka\cdot\bold{x}} \frac{\cosh(\mu k(1+H(\bold{x})))}{\cosh^2(\mu k)}\frac{\ka}{k^2}\right].
\end{equation}
In this equation $\nabla = (\partial_x,\partial_y)$. Equation (\ref{S3:Eq02}) couples the topographic 
coefficients $X_{\ka}$ with the Dirichlet data, where  the topography profile $H(\x)$ appears in the relation.

Assuming that the coefficients  $X_\ka$ have been computed by means of equation (\ref{S3:Eq02}), we can use the representation formula (\ref{S3:Eq01}) to compute the vertical derivative of the velocity potential at $z=0$, 
thus characterizing the Dirichlet-to-Neumann operator acting on $q$:
\begin{equation}\label{S3:Eq03}
G[q] = \phi_z(\x,0) = \sum_{\ka\in\Lambda^*}e^{i\ka\cdot\x}\left[\hat{q}(\ka)\mu k\tanh(\mu k) + X_\ka\mu\sech^2(\mu k)\right].
\end{equation}
Notice that the presence of a non trivial topography $H$ yields an extra term in the Fourier representation of the operator $G$, 
when we compare (\ref{S3:Eq03}) with (\ref{S2:Eq11}).  We have still to describe how to compute the topographic coefficients
 $X_\ka$.


We define the following linear operators, acting on the space of bounded sequences $\ell^{\infty}(\Lambda^*)$ as follows: for any $f_\ka\in\ell^{\infty}(\Lambda^*)$, 
\begin{eqnarray} 
A[f_\ka] &=& \sum_{\ka\in\Lambda^*} f_\ka\nabla \cdot \left[ e^{i\ka\cdot\bold{x}} \frac{\sinh(\mu k H(\bold{x}))}{\cosh(\mu k)}\frac{\ka}{k}\right], \label{S3:Eq04}\\ 
B[f_\ka] &=& \sum_{\ka\in\Lambda^*} f_\ka\nabla\cdot \left[ e^{i\ka\cdot\bold{x}} \frac{\cosh(\mu k(1+H(\bold{x})))}{\cosh^2(\mu k)}\frac{\ka}{k^2}\right].\label{S3:Eq05}
\end{eqnarray}
By introducing these linear operators, equation (\ref{S3:Eq02}) becomes
\begin{equation}\label{S3:Eq06}
B[X_\ka]=A[\hat{q}(\ka)].
\end{equation}
As a consistency test we compute these operators in the flat bottom case $H = 0$. 
Then  for any bounded sequence of complex numbers $f_\ka$, $A[f_\ka]$ vanishes because of the $\sinh$ function in the numerator, whereas for the operator $B$ we have 
\begin{equation}
B[f_\ka] = \sum_{\ka\in\Lambda^*} f_\ka\nabla\cdot \left[ e^{i\ka\cdot\bold{x}} \frac{\cosh(\mu k)}{\cosh^2(\mu k)}\frac{\ka}{k^2}\right] = \sum_{\ka\in\Lambda^*}e^{i\ka\cdot\bold{x}}\sech(\mu k)f_\ka.
\end{equation}
So in the case $H=0$, equation (\ref{S3:Eq06}) simplifies to
\begin{equation}
B[X_\ka] = \sum_{\ka\in\Lambda^*}e^{i\ka\cdot\bold{x}}\sech(\mu k)X_\ka = 0.
\end{equation}
This is only possible if the coefficients $X_\ka$ vanish because of the orthogonality of the complex exponential functions. Thus the operator given in (\ref{S3:Eq03}) reduces to the classical flat bottom operator $G$ in (\ref{S2:Eq11}).

What happens when the depth is constant ($H=h_0$) but different from the reference depth? 
We show that the operator defined through expressions (\ref{S3:Eq02}) and (\ref{S3:Eq03}) is the DtN 
operator corresponding to the constant depth $1+h_0$. This gives us an idea for the vertical flow structure over
a locally flat region.

Assuming that $H = h_0$ is constant, and $f_\ka$ is a bounded sequence of complex numbers,
\begin{equation}
A[f_\ka] = \sum_{\ka\in\Lambda^*} f_\ka\nabla \cdot \left[ e^{i\ka\cdot\bold{x}} \frac{\sinh(\mu k h_0)}{\cosh(\mu k)}\frac{\ka}{k}\right] = \sum_{\ka\in\Lambda^*} e^{i\ka\cdot\bold{x}} k\frac{\sinh(\mu k h_0)}{\cosh(\mu k)} f_\ka.
\end{equation}
On the other hand we have that
\begin{equation}
B[f_\ka] = \sum_{\ka\in\Lambda^*} f_\ka\nabla \cdot \left[ e^{i\ka\cdot\bold{x}} \frac{\cosh(\mu k (1+h_0))}{\cosh^2(\mu k)}\frac{\ka}{k^2}\right] = \sum_{\ka\in\Lambda^*} e^{i\ka\cdot\bold{x}} \frac{\cosh(\mu k (1+h_0))}{\cosh^2(\mu k)} f_\ka.
\end{equation}
Equation (\ref{S3:Eq06}) now yields the following relation between the known Fourier coefficients $\hat{q}(\ka)$ and the unknown coefficients $X_\ka$:
\begin{equation}
B[X_\ka] = \sum_{\ka\in\Lambda^*} e^{i\ka\cdot\bold{x}} \frac{\cosh(\mu k (1+h_0))}{\cosh^2(\mu k)} X_\ka = \sum_{\ka\in\Lambda^*} e^{i\ka\cdot\bold{x}} k\frac{\sinh(\mu k h_0)}{\cosh(\mu k)} \hat{q}(\ka) = A[\hat{q}(\ka)].
\end{equation}
This equation determines $X_\ka$ uniquely in terms of $\hat{q}(\ka)$ because of the orthogonality of the basis functions. The solution is:
\begin{equation}
X_\ka = \hat{q}(\ka)k\cosh^2(\mu k)\left(\tanh(\mu k (1+h0))-\tanh(\mu k )\right).
\end{equation}
Finally equation (\ref{S3:Eq03}) yields the consistent result that
\begin{equation}
G[q] = \sum_{\ka\in\Lambda^*}\hat{q}(\ka)\mu k\tanh(\mu k (1+h0)).
\end{equation}

From the point of view of functional analysis the linear operator $B$ is a Hilbert-Schmidt operator whose kernel is given by: 
\begin{equation}
\nabla\cdot \left[ e^{i\ka\cdot\bold{x}} \frac{\cosh(\mu k(1+H(\bold{x})))}{\cosh^2(\mu k)}\frac{\ka}{k^2}\right].
\end{equation}
Such operators are limits of finite rank operators (\cite{DunfordSchwartz}) and therefore they are compact operators.
An important result due to \citet{CSNG}, regards the solvability of equation (\ref{S3:Eq06}): 
the operator $B$ is always invertible and thus, in principle, it is always possible to solve for the coefficients $X_\ka$ 
in terms of $\hat{q}(\ka)$.  
On the other hand inverting $B$, whose range in nearly finite dimensional, must be done with care 
in order to avoid ill-conditioned behaviour that may arise from the 
discrete versions of the problem. 

Some ill-conditioned problems, that arise from this kind of operators, have been a matter of research in recent years, particularly in the work of 
\citet{OliverasDeconinck} and \citet{VasanDeconinck} through the AFM formulation. 
More recently, a very detailed numerical study was carried out by \citet{WilkeningVasan} comparing different DtN formulations (including AFM and AFM$^\ast$).

Under this perspective, the action of the Dirichlet-to-Neumann operator will be examined in the presence of
highly variable topographies. In the next section, we present accurate computational results 
for unexplored regimes of the bottom topography.   
But first, we describe our pseudo-spectral numerical method, based on a (physically motivated) Galerkin approximation
for the topographic components of the DtN operator.

\subsection{The Galerkin approximation}
We now introduce a Galerkin approximation for equation (\ref{S3:Eq02}). 
We have a compact linear operator
\begin{equation}
B:\ell^2(\Lambda^*) \longrightarrow L^2(T^2),
\end{equation}
together with a linear problem of the form
\begin{equation}
B[X_\ka] = Y,
\end{equation}
where $Y = A[\hat{q}(\ka)]$ is a continuous function of the space variable $\x$.

Let $e_\ka\in\ell^2(\Lambda^*)$ be the canonical vector. For $M>0$, let $U_M = span\{\delta_\ka \mid k\leq M\}\subset \ell^2(\Lambda^*)$.  Let $V_\M = span\{e^{i\ka\cdot\x} \mid \ k\leq M\}$, be the corresponding linear subspace of $L^2(T^2)$. Notice that both finite dimensional vector spaces have the same dimension.

For any sequence of complex numbers $X_\ka\in U_M$, the action of the operator $B$ is given by the truncated sum:
\begin{equation}\label{S3:Eq20}
B[X_\ka] = \sum_{k\leq M}X_\ka\nabla\cdot \left[ e^{i\ka\cdot\bold{x}} \frac{\cosh(\mu k(1+H(\bold{x})))}{\cosh^2(\mu k)}\frac{\ka}{k^2}\right].
\end{equation}

Although the right hand side of (\ref{S3:Eq20}) defines a differentiable function, the spectral content of such function is unknown. 
If we consider momentarily the case where $H = 0$, then the right hand side of (\ref{S3:Eq20}) defines a band limited function. However due to variations of the bottom geometry $H=H(\x)$, the spectral content, of the otherwise band limited function, is spread over the Fourier spectrum in $L^2(T^2)$. 

According to the truncated expression  (\ref{S3:Eq20}) define $R_M \equiv B[X_\ka] - Y$ as the residual, for a 
given value of the Galerkin parameter. We will find conditions for the residual to be orthogonal to the subspace $V_M\subset L^2(T^2)$.
We first compute the inner product of the residual  with one of the basic functions $e^{i\ele\cdot\x}\in V_M$.
\begin{equation}\label{S3:Eq21}
\left(R_M,e^{i\ele\cdot\x}\right) = \left(B[X_\ka] - Y,e^{i\ele\cdot\x}\right) = \left(B[X_\ka],e^{i\ele\cdot\x}\right) -\left(Y,e^{i\ele\cdot\x}\right).
\end{equation}

A necessary and sufficient condition for the orthogonality of the reminder function is that for $e^{i\ele\cdot\x}\in V_M$, 
\begin{equation}\label{S3:Eq22}
\left(B[X_\ka],e^{i\ele\cdot\x}\right) = \left(Y,e^{i\ele\cdot\x}\right).
\end{equation}

The right hand side of (\ref{S3:Eq22}) corresponds to the Fourier coefficient of the function $Y$, denoted by $\hat{Y}(\ele)$. Then equation (\ref{S3:Eq22}) yields the following square system of linear equations:
\begin{equation}\label{S3:Eq23}
\begin{aligned}
\left(B[X_\ka],e^{i\ele\cdot\x}\right) &= \int_{T^2}e^{-i\ele\cdot\x}\sum_{k\leq M}X_\ka\nabla\cdot \left[ e^{i\ka\cdot\bold{x}} \frac{\cosh(\mu k(1+H(\bold{x})))}{\cosh^2(\mu k)}\frac{\ka}{k^2}\right]\ d\x\\ 
&= \sum_{k\leq M} i(\ele\cdot\ka) X_\ka\int_{T^2} e^{-i(\ele-\ka)\cdot\x}\left[\frac{\cosh(\mu k(1+H(\x)))}{k^2\cosh^2(\mu k)}\right]\ d\x = \hat{Y}(\ele).
\end{aligned}
\end{equation}

Let $\tilde{X}_\ka\in U_M$ be a solution to the system of linear equations (\ref{S3:Eq23}). 
In order to compute the action of the Dirichlet-to-Neumann operator on a given input function $q(\x)$ we proceed as follows:
(i) the first term in equations (\ref{S3:Eq03}) is easily computed with an FFT; 
(ii) for the second term, we fix a value of the parameter $M$ and compute an approximate solution $\tilde{X}_\ka$ 
through system (\ref{S3:Eq23}). This approximate solution is zero-padded and  plugged into the second term of equation (\ref{S3:Eq03}). These steps yield an approximation for the Neumann data $G[q]$. 

As noted, the ``flat part" of the DtN operator (\ref{S3:Eq03}) is straightforward and stable numerically. We use all Fourier modes available on our  mesh. In order to find the topographic coefficients
of the DtN's  Fourier decomposition  (\ref{S3:Eq03}) the linear system (\ref{S3:Eq23})
has to be inverted. 
The dimension of the respective subspace of interest can be kept small. 

Next we present 
a physically motivated  strategy to truncate to topographic component of the DtN operator, namely keeping only the relevant modes
that do interact with the topography. This guides our choice for the Galerkin parameter $M$. 
Higher (short-wave) modes which do not interact with the topography are discarded by the projection onto the lower
dimensional subspace determined by the Galerkin method. The presence of higher modes (in a deep water
regime) would only contribute to the ill-conditioning of system  (\ref{S3:Eq23}).

\subsubsection{A physically motivated choice for the  Galerkin parameter $M$.}

As mentioned above, the Galerkin parameter $M$ plays an important role in the simulations.  
It can be chosen independently of the number of Fourier modes available on our mesh and it controls 
the size of the linear system to be inverted.  Therefore it has an impact on the numerical performance of the DtN operator.

The choice of the Galerkin parameter comes from  physical properties related to the water wave problem. 
Namely related to  particle trajectories underneath the surface waves. 

Lets assume momentarily that we are in the flat bottom case. The fluid particle trajectories, beneath a time harmonic wave, 
describe approximately an elliptical orbit. Let:
\begin{equation*}
\begin{aligned}
\eta(x,t) &= Ae^{ikx-\omega t},\\
\phi(x,0,t) &= c + \tfrac{iA}{\omega}e^{ikx-\omega t},
\end{aligned}
\end{equation*}
be the time harmonic free surface elevation and its corresponding velocity potential. 
Then a classical result, found in \cite{Constantin} and \cite{DeanDalrymple}, shows that at depth $z_0$
the semi-axis of the ellipse are given by:
\begin{equation*}
\begin{aligned}
\text{major semi-axis}: &\ \frac{A\cosh(\mu k(z_0+1))}{\sinh(\mu k)},\\
\text{minor semi-axis}: &\ \frac{\mu A\sinh(\mu k(z_0+1))}{\sinh(\mu k)}.
\end{aligned}
\end{equation*}
In order to choose the Galerkin parameter $M$ we set $z_0 = \inf_x H(x)$. For a 
given tolerance $\delta>0$, we let $k = k^\ast$ be the smallest wavenumber such that:
\begin{equation*}
\frac{\cosh(\mu k^\ast (z_0+1))}{\sinh(\mu k^\ast)} \leq \delta.
\end{equation*}
We set $M = k^\ast$. 
Particle orbits due to higher modes ($k > k^\ast$) are very small near the topography.  
Hence these short waves do not contribute to the topographic coefficients.
It is important to notice that our physical choice of the Galerkin parameter $M$ depends only on $\mu$, $H$, and $k^\ast$. 
This procedure selects the wavenumber band whose interaction with the bottom is nontrivial and uses this smaller subspace 
in the computation of the topographical contribution to the Dirichlet-to-Neumann operator. We will exhibit 
accurate simulations in the presence of  highly nontrivial topographies.

In most of our simulations the large amplitude topography reaches half the depth. 
Therefore we have chosen $z_0 = 0.5$ in finding $M$.  
The dispersion parameter is $\mu = 0.1$  and  our tolerance for particle motion was $\delta = 10^{-5}$. 
The corresponding value of the Galerkin parameter is $M = 230.25$. 
For a periodic domain of (dimensionless) length equal to $10$, this implies that only the first $366$ 
Fourier modes are needed for constructing the topographic component of the DtN operator. 
The typical mesh size considered in our simulations,
per side of the free surface, 
ranges from $2^{9}$ to $2^{11}$. The advantage of choosing an appropriate,
physically motivated $M$ is clear.


\subsection{The DtN operator through conformal mapping}

In the presence of a non trivial bottom geometry, the elliptic boundary value problem (\ref{S3:Eq02}) - (\ref{S3:Eq04}) has no closed form analytical solution. 
In two dimensions it is very efficient to define the DtN operator through the conformal mapping of the variable-depth channel.
But this strategy does not apply in three dimensions. 
Nevertheless in order to benchmark our present method we will first compare two dimensional solutions with those
obtained by a conformal mapping technique  (\citet{Nachbin2003}) which we here briefly summarize. 

The cornerstone of the method is finding a conformal mapping from a uniform strip, in the $(\xi,\zeta)-$plane, onto the 
variable-depth fluid domain in the $(x,z)-$plane. In the physical domain the level-curves 
 $\xi$-constant and $\zeta$-constant generate an orthogonal curvilinear coordinate system. In this new coordinate 
 system the Laplacian remains invariant and the bottom boundary is flatten out, being a level-curve in $\zeta$. 
 In general the determination of the conformal mapping is a difficult task. However in the case of polygonal topographies 
 this is achieved with the Schwarz-Christoffel Toolbox (\citet{Driscoll2002,FokasNachbin}).

Having the conformal mapping at hand, the elliptic boundary value problem is written 
in the curvilinear coordinate system $(\xi,\zeta)$as:
\begin{eqnarray}
\mu^2\phi_{\xi\xi} + \phi_{\zeta\zeta} &=& 0, \ \ \ \text{in $-1<\zeta<0$,}\label{S3:Eq24}\\
\phi(\xi,0) &=& q(\xi),\ \ \ \text{on $\zeta = 0$.}\label{S3:Eq26}\\
\phi_\zeta &=& 0, \ \ \ \text{on $\zeta = -1$,}\label{S3:Eq25}
\end{eqnarray}

In this setting it is very easy to compute $\phi_{\zeta}(\xi,0)$ by means of formula (\ref{S2:Eq11}). 
We then proceed to compute $\phi_z(x,0)$ from $\phi_{\zeta}(\xi,0)$ using the following formula (\cite{Nachbin2003}):
\begin{equation}\label{S3:Eq27}
\phi_z(x(\xi,0),0)  = \frac{\phi_{\zeta}(\xi,0)}{M(\xi)}.
\end{equation}
The variable coefficient $M(\xi)$ is defined as
\begin{equation}\label{S3:Eq28}
M(\xi)  = z_\zeta(\xi,0),
\end{equation}
and is the square root of the Jacobian evaluated at the undisturbed free surface. Hence a smooth coefficient.
This metric coefficient is easily computed using the Schwarz-Christoffel Toolbox  as described in \citet{FokasNachbin}.

\begin{figure}
\centerline{\includegraphics[width = \textwidth]{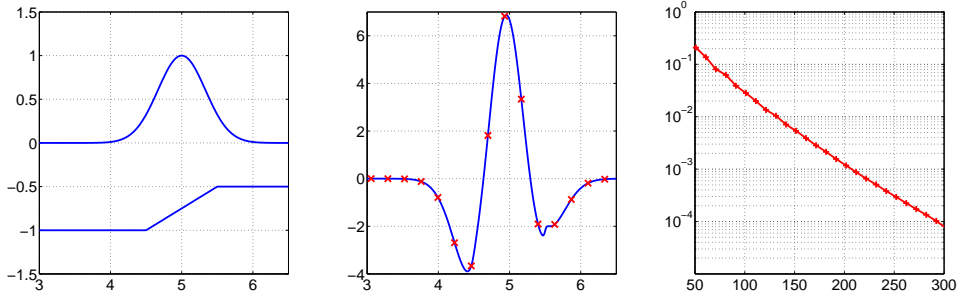}}
\caption{Left Panel: Gaussian potential along the undisturbed free surface  ($z=0$) over a large amplitude sloping bottom
located at $(4.5\leq x\leq5.5; -1\leq z \leq -0.5)$. Central Panel: 
The respective Neumann data $\phi_z(x,0)$ computed through the DtN operator acting on the Gaussian potential ($\mu = 0.05$). The 
solid curve was obtain through equations (\ref{S3:Eq02}) and (\ref{S3:Eq03}), while the red crosses were computed with the conformal mapping technique (\ref{S3:Eq27}). Both methods used $2^{10}$ points. Right Panel (log-plot): Relative error between the Galerkin approximation and the 
conformal mapping solution,  as a function of the Galerkin parameter $M$. }
\label{Fig01}
\end{figure}

In order to compare both methods we start with the time independent example as given by (\ref{S3:Eq24})-(\ref{S3:Eq25}). 
As our Dirichlet data, we consider a Gaussian pulse located above the sloping bottom, as 
depicted at the left of figure \ref{Fig01}. The bottom topography is not smooth.
At the central panel of figure \ref{Fig01} the solution with the present nonlocal method (solid line) is in very 
good agreement with that computed through the conformal mapping method (crosses). 
At the right panel of figure \ref{Fig01} we perform a resolution study for the Galerkin approximations, as the parameter $M$ increases. 
The solid line represents the relative error between the Galerkin approximation and the solution with the conformal mapping technique. 
For example  when $M = 300$ the relative error is $0.89107\times10^{-3}$. 

\section{Water wave simulations}

We now present numerical results for the time evolution of surface water waves propagating over
highly variable topographies.  
The time evolution has been performed with the fourth-order Runge-Kutta method. 

We use the following initial data along the free surface $z=0$:
\begin{equation}
\phi(x,0,0)  = e^{-({(x-x_0)}/{\sigma})^2},
\label{phi0}
\end{equation}
is the initial velocity potential. We use $\sigma = 1/6$ so that the pulse is approximately of unit width,
centered at $x=x_0$; the initial wave elevation is 
\begin{equation}
 \eta(x,0) = F^{-1}\left[\frac{k tanh(\mu k)}{\mu}F[\phi]\right].
 \label{eta0}
 \end{equation}
 By $F$ we indicate the use of an FFT and its inverse $F^{-1}$ accordingly. 
 
 In the three dimensional simulations we have a plane wave with the above profile.
 
\subsection{Two dimensional results}

 We start with two-dimensional simulations where results are known and we can further compare with the conformal mapping 
technique.

\subsection{Waves over a submerged structure}
We again consider a non-smooth topography, now in the form of a large submerged trapezoidal structure. 
 In the  conformal mapping technique the Jacobian is a 
 smooth function along the free surface and therefore the corners of the submerged structure have been regularized.  In the
 present method the topography comes into the formulation in its original form.
 
\begin{figure}
\centerline{\includegraphics[width = \textwidth]{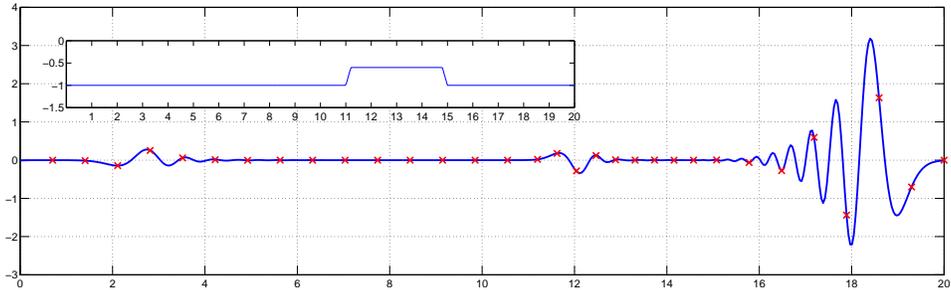}}
\caption{Evolution of the wave  elevation after $12$ units of time. 
The solution with the present DtN method is displayed by a solid line while with crosses  for the conformal mapping technique. 
For a better visualization, only a subset of the crosses were depicted.
The wave regime corresponds to $\mu = 0.05$. The inset shows the topography at $11<x<15$ with an amplitude of 0.5, half the depth.}
\label{Fig02}
\end{figure}

 A Gaussian pulse of unit width for $\phi$, (similar to that of figure \ref{Fig01}) propagates from the left ($x<10$) towards the submerged
 structure located at $11<x<15$ (see inset of figure \ref{Fig02}). 
The conformal mapping method follows the work presented in  \citet{NachbinArtiles}.
  A snapshot of the wave elevation at time $t=12$ is shown in figure \ref{Fig02}. The agreement
between the two methods is very good, regarding both the dispersive wave profile as well as phase speed. 
The solid line depicts the solution with the present method while the crosses depict  results with the conformal mapping
technique. 
Also the two 
reflected waves (due to climbing and descending from the structure) are accordingly captured by the methods.
For both simulations we used $2^{10}$ points in physical space, and we considered all the Fourier modes available in our computational grid. 

%
%
We should also mention that we found a  maximum relative error of $6.275\times10^{-4}$ confirming the 
very good agreement between the two methods in this example. 

\subsection{Periodic topographies}

In this section we study  a wave moving from a region of constant depth towards a periodic patch of
large amplitude depth variations. 
This is a problem with known mathematical results and is a good test for our DtN formulation, in particular the calculation of its topographic
coefficients with our Galerkin approach. 
We consider mean-zero  periodic depth variations with an amplitude as high as half of the channel depth. 
The wave-topography interaction is considered in two different regimes.
First the topographic variations are on the same scale as the wave field. Then we allow for large amplitude, rapidly varying depth variations,
a nontrivial test for the present DtN formulation. In this regime the bottom slope is large, while the wave feels the topography in 
an averaged form, as reported by  \cite{RosalesPapanicolaou}.

From now on only our numerical solution will be presented.

\subsubsection{Bragg resonance}

In the context of water waves Bragg resonance arises when the wavelength is twice as long as the topography's period.
The reflection is maximum in this regime. 
This is a known problem in water waves and to provide a few references we mention the work of \cite{Mei1985}, regarding linear waves, and 
 the more recent work of \cite{choiMilewski}, regarding nonlinear waves. 
The setting for our  simulations differs from the theoretical work of \citet{Mei1985} and \citet{choiMilewski}. 
The topographies considered in the theoretical investigations had small amplitudes of order of $\mathcal{O}(\mu^2)$. 
In the present simulations the channel's depth oscillates from $0.5$ to $1.5$.

\begin{figure}
\centerline{\includegraphics[width = \textwidth]{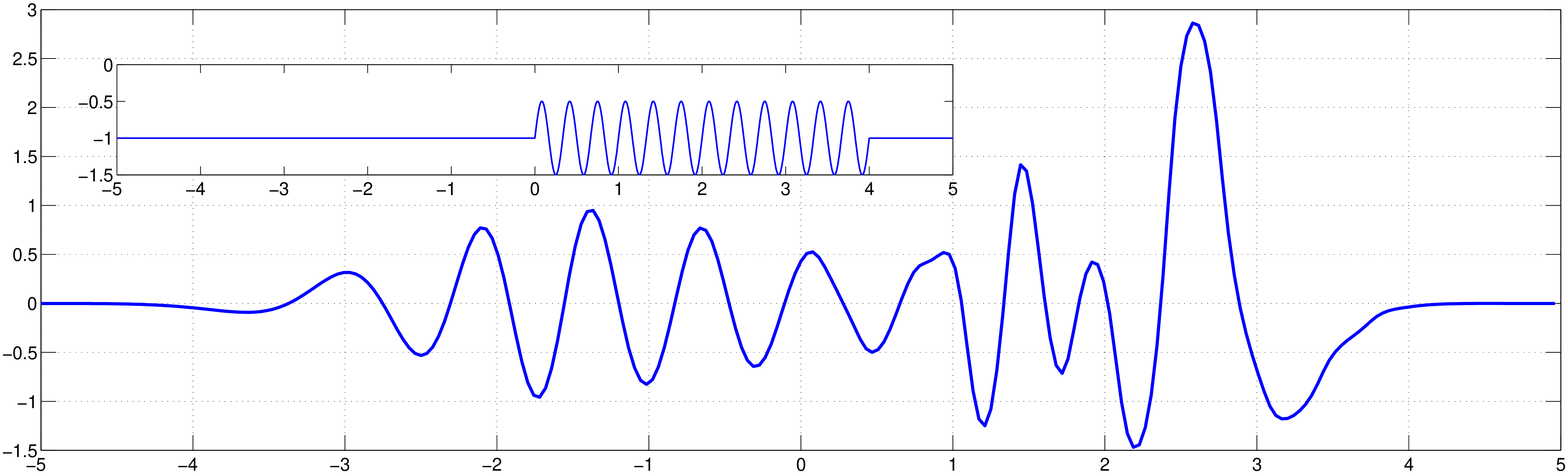}}
\caption{Evolution of the wave elevation after $4.5$ units of time. Notice the nearly monochromatic structure of the reflected wave. The wave regime corresponds to $\mu = 0.1$. The inset shows the underwater sinusoidal topography. The numerical method used $2^{10}$ Fourier modes
and the Galerkin method used $M=250$ for the topographic component of the DtN operator.}
\label{Fig03}
\end{figure}

In the figure \ref{Fig03} we present a snapshot of the solution at time $t=4.5$. A  pulse propagates to the right, from the constant
depth region located at $x<0$ (see figure \ref{Fig03}). As the wave propagates over the periodic patch a 
(nearly) monochromatic reflected wave
is observed (moving to the left) with a wavelength twice as long as the topography's period. 
The observed reflected wavelength is $0.667$ (for $-2<x<0$),  where the  sinusoidal
topography has period  $0.333$.

\subsubsection{Homogenization with a rapidly varying topography}

Now we investigate rapidly varying periodic topographies. This is an important test to see if our DtN formulation
accurately captures fine features of the topography.  The topography has large slopes and we want to observe numerically
the impact of our Galerkin approximation in handling the  topographic component of the DtN operator.

Through a multiscale analysis \citet{RosalesPapanicolaou} showed that a long wave feels the fine features 
of the topography in an averaged form. Namely a homogenization effect takes place to leading order. 
The wave behaves as if propagating over an ``effectively flat" region:  no reflection is observed and the wave speed
is smaller than the (dimensionless) unit speed. The periodic topography is mean-zero but the wave behaves as
if there was an effective depth smaller than 1. 
This was found by \citet{RosalesPapanicolaou} in the weakly nonlinear, weakly dispersive case. 
The same result holds  in the linear case, as reported in \citet{Nachbin1993}.
The non-local formulation has been used in theoretical investigations of this phenomena. \cite{CSNG} derived an asymptotically valid KdV equation from the non-local equations. Their KdV equation also has a smaller effective speed of propagation.
Our simulations consider a large amplitude, rapidly varying topography. 
As before, our topography varies from $0.5$ to $1.5$ and the sinusoidal bottom oscillates $12$ times per (unit) pulse width.

\begin{figure}
\centerline{\includegraphics[width = \textwidth]{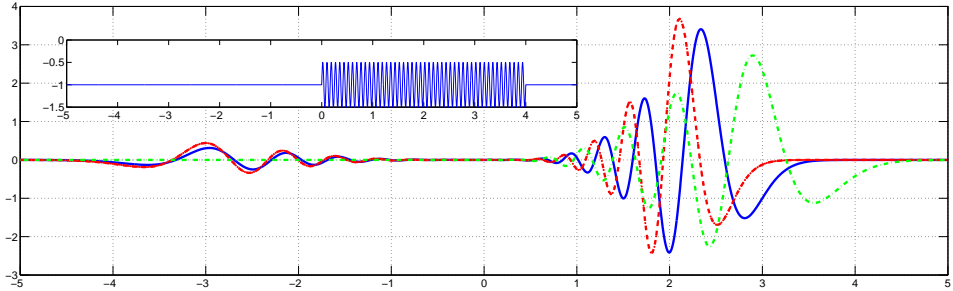}}
\caption{Solid line: Homogenised wave field. Dashed dotted line: Simulation without the topography. Dotted line: Simulation over the envelope of the bottom. All the simulations correspond to the regime $\mu = 0.1$.
The numerical method used $2^{11}$ Fourier modes
and the Galerkin method used $M=250$ for the topographic component of the DtN operator.}
\label{Fig04}
\end{figure}

In figure \ref{Fig04} we superimpose three simulations: (a) a flat bottom case; (b) a large amplitude, zero-mean, rapidly varying
periodic topography; (c) a step obtained as the envelope of the rapidly varying topography. The dispersion parameter is
taken as $\mu=0.1$. We should recall  that the Galerkin parameter $M$ depends on $\mu$. The shallow water regime is easier to solve than a dispersive regime. The tests reported below were also done for a (nearly) traveling pulse in a very shallow
channel, but we only report the more demanding dispersive case.
 
The (green) dotted line corresponds  to the flat bottom case. No reflected wave is observed. At the wavefront we
notice the dispersive oscillatory tail.
We focus on the speed at the wave front, which in this case is $0.94$ due the initial wave profile and
dispersion.
The (red) dashed-dotted line corresponds to the envelope of the bottom topography, namely a step. 
The wave front appears behind the flat bottom case because over the step the speed of propagation is smaller, about $0.68$.
Finally the (blue) solid line is the solution in the presence of the rapidly varying topography.
The effective speed is smaller than the flat bottom case but larger than the step. This is our numerically
homogenized wave field. 
No reflection is produced over the rapidly varying periodic topography.
The bottom's fine features where captured accordingly along the free surface through the
DtN operator. The solution is free from numerical noise indicating that the Galerkin approximation for the 
topographic coefficients of the non-local operator performed well.  
The numerically observed  (effective) wave speed is of $0.82$. 
In the shallow water case we clearly see a (nearly) traveling wave over the topography, with no reflected wave observed. 
The only reflected signal comes from the transition from flat to periodic.


%

\subsection{The three dimensional  DtN operator: refraction by a submerged mound}

In this section we investigate the refraction of two dimensional surface waves due to bathymetric variations. 
We explore with  changes in  propagation direction as the wave  passes over a circular submerged mound. 
We will report on a special mound, known as the Luneburg lens,  where the front of an incoming plane wave 
bends in a particular fashion.

\subsubsection{The Luneburg lens}

We consider a particular submerged mound which, as we shall see, plays a role analogous  to that of the Luneburg lens in optics. 
The mound focuses the wave energy at a determined point in space. 
Our goal is to show that the Dirichlet-to-Neumann operator, in our two-dimensional surface model,  accurately captures 
the complex refraction pattern that arises due to the non-trivial change in depth.

The underwater mound is given by 
\begin{equation}\label{S5:Eq01}
H(r) = 
\begin{cases}
\frac{\alpha^2}{\alpha^2 + 1 - ({r}/{r_0})^2} - 1, \ \ \ \text{if $r < r_0,$}\\
0, \ \ \ \text{if $r\geq r_0$.}
\end{cases}
\end{equation}
We consider $\alpha = 0.8$, $r = \sqrt{(x-8)^2+(y-10)^2}$ and $r_0 = 4$. The water depth above the center of the mound is about 
$0.60$.
The numerical method used $2^{9}$ Fourier modes in each direction
and the Galerkin method used $M=37.69$ for the topographic component of the DtN operator. The use of a smaller $M$ is 
due to memory restrictions. For a two dimensional surface and $M=37.69$ we have $240^2$ Fourier modes representing
functions in our respective subspace for the topographic component of DtN. In order to find the contribution for 
each of these Fourier modes we need to build and invert a matrix $240^2$ by $240^2$.

\begin{figure}
\centerline{\includegraphics[width = \textwidth,keepaspectratio]{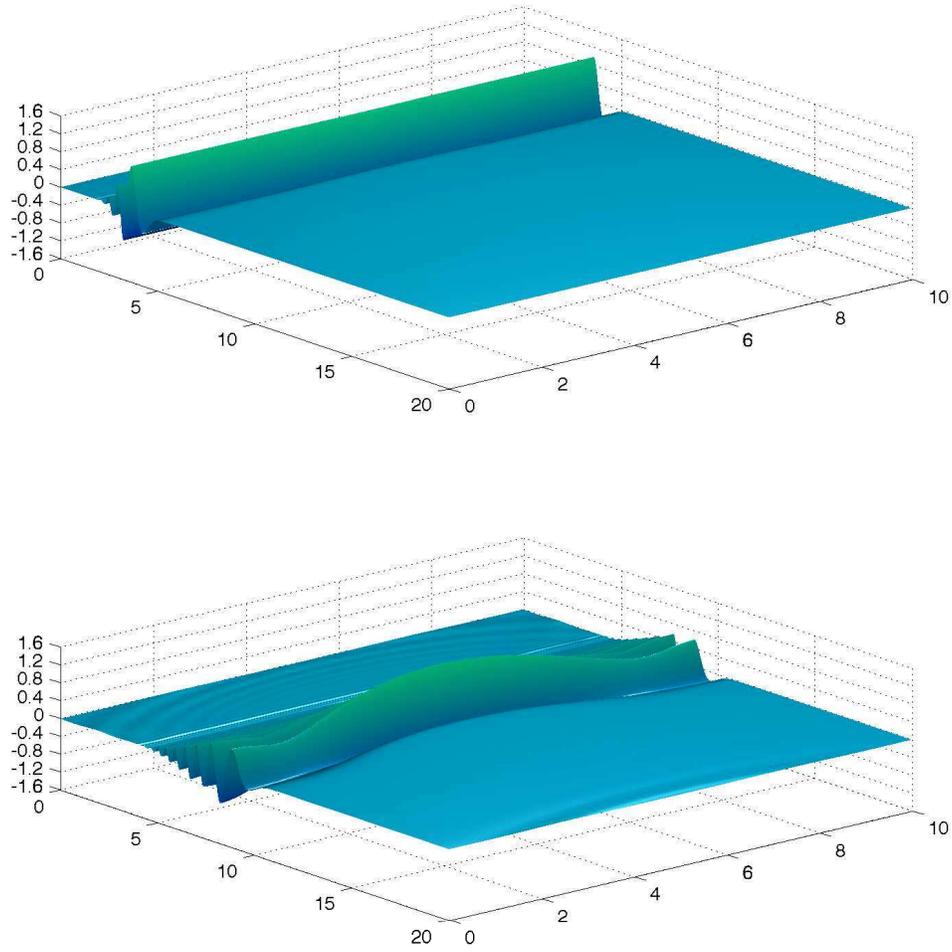}}
\caption{Snapshots  at times $t = 3.1$ and $t=8.6$.  The incoming surface elevation is propagating towards the Luneburg lens 
($\alpha = 0.8$)  in the $\mu = 0.1$ regime. An animation for the corresponding velocity potential can be found in the supplemental material of this article.}
\label{Fig05}
\end{figure}
\begin{figure}
\centerline{\includegraphics[width = \textwidth]{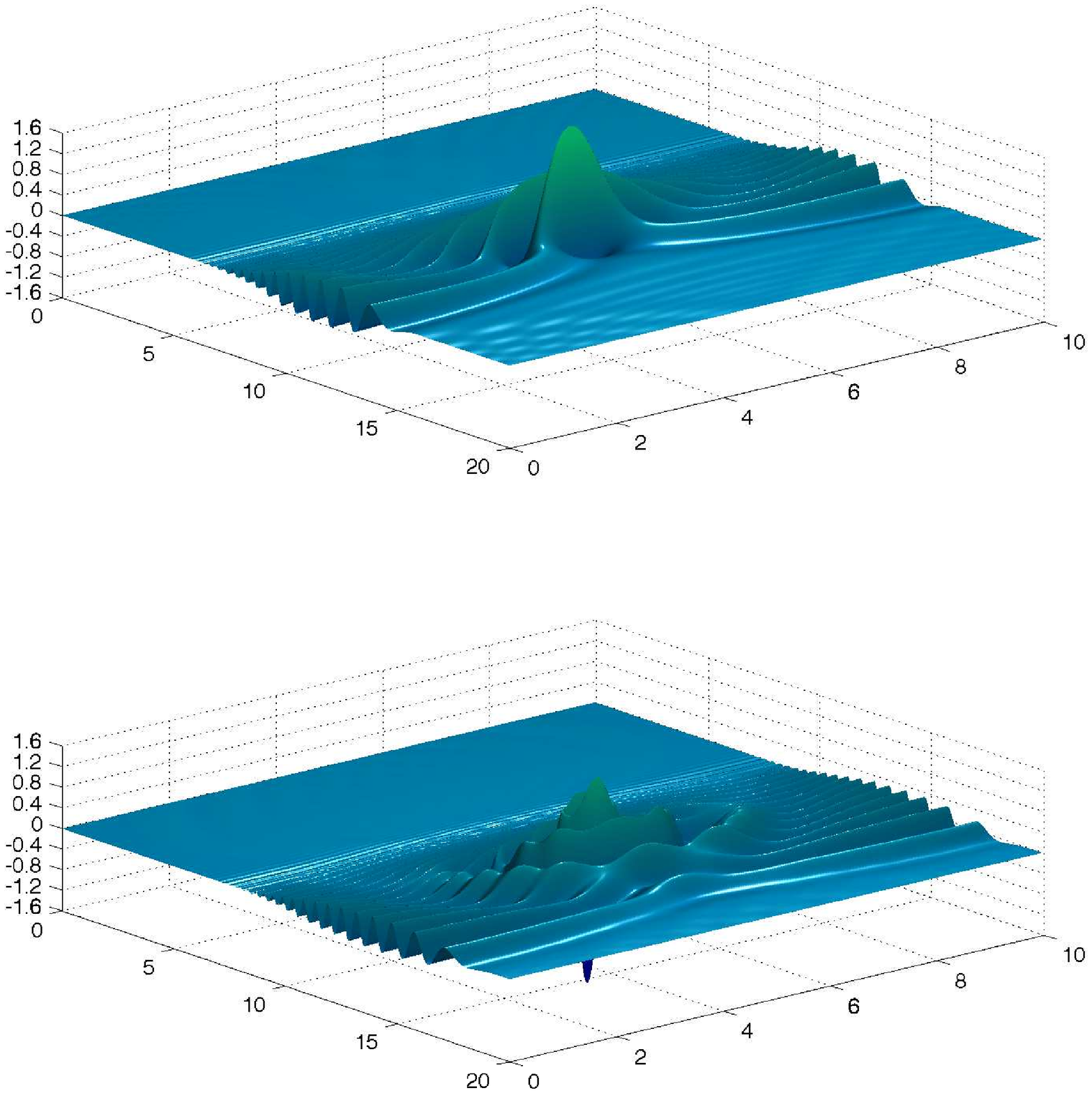}}
\caption{
Snapshots  at times $t = 13.6$ and $t=17$.  The  surface elevation is propagating over the Luneburg lens 
($\alpha = 0.8$)  in the $\mu = 0.1$ regime. The focusing due to refraction is clearly observed.}
\label{Fig06}
\end{figure}

Four snapshots of the solution are presented in  figure \ref{Fig05} and figure \ref{Fig06}. In the top 
panel of figure \ref{Fig05}, corresponding to time $t = 3.1$, we see the incoming plane wave (from the left)
already displaying some dispersive effect. We have used $\mu = 0.1$.
A short oscillatory tail is apparent behind the wave front. 
The topography is in the middle of the computational domain and  no change of direction of propagation is observed.

In the lower panel of figure \ref{Fig05}, at time $t = 8.6$, we see a strong deformation along the wavefront  due to its interaction with the 
submerged mound. The depth in the middle is decreasing and therefore the speed at the central part of wavefront is smaller.

At the top of figure \ref{Fig06} we have the solution at time $t = 13.6$. We observe the effect of the Luneburg lens: 
it focuses the wave energy at a point behind the mound. 
At this time the largest wave amplitude is observed, exactly at the focusing point as will be confirmed below. 
The wave amplitude at the focusing point is about twice as high as that of the initial incoming wave.
As the high amplitude peak moves down due to gravity  smaller waves (of the dispersive tail) will focus behind it, 
giving rise to a circular pattern that eventually dominates the wave field, at time $t = 17$ (bottom of figure \ref{Fig06}).
We tested with other shapes of the submerged mound (such as a Gaussian profiles) and of course the focusing
effect does not take place.

\begin{figure}
\centerline{\includegraphics[width = \textwidth]{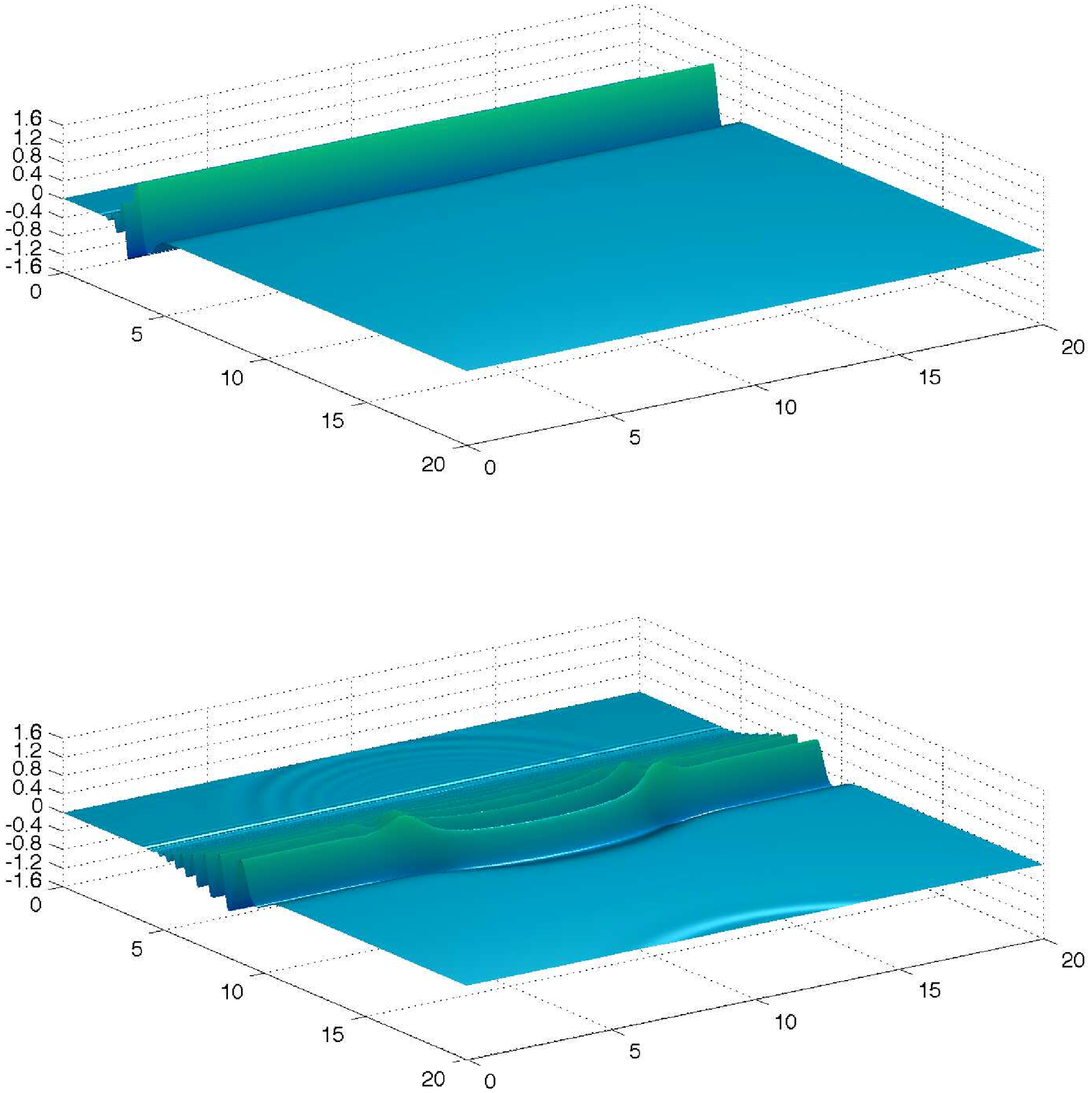}}
\caption{
Snapshots  at times $t = 3.1$ and $t=8.6$.  The incoming surface elevation is propagating towards a concave cavity 
(with the Luneburg lens profile; $\alpha = 0.8$)  in the $\mu = 0.1$ regime. }
\label{Fig07}
\end{figure}
\begin{figure}
\centerline{\includegraphics[width = \textwidth]{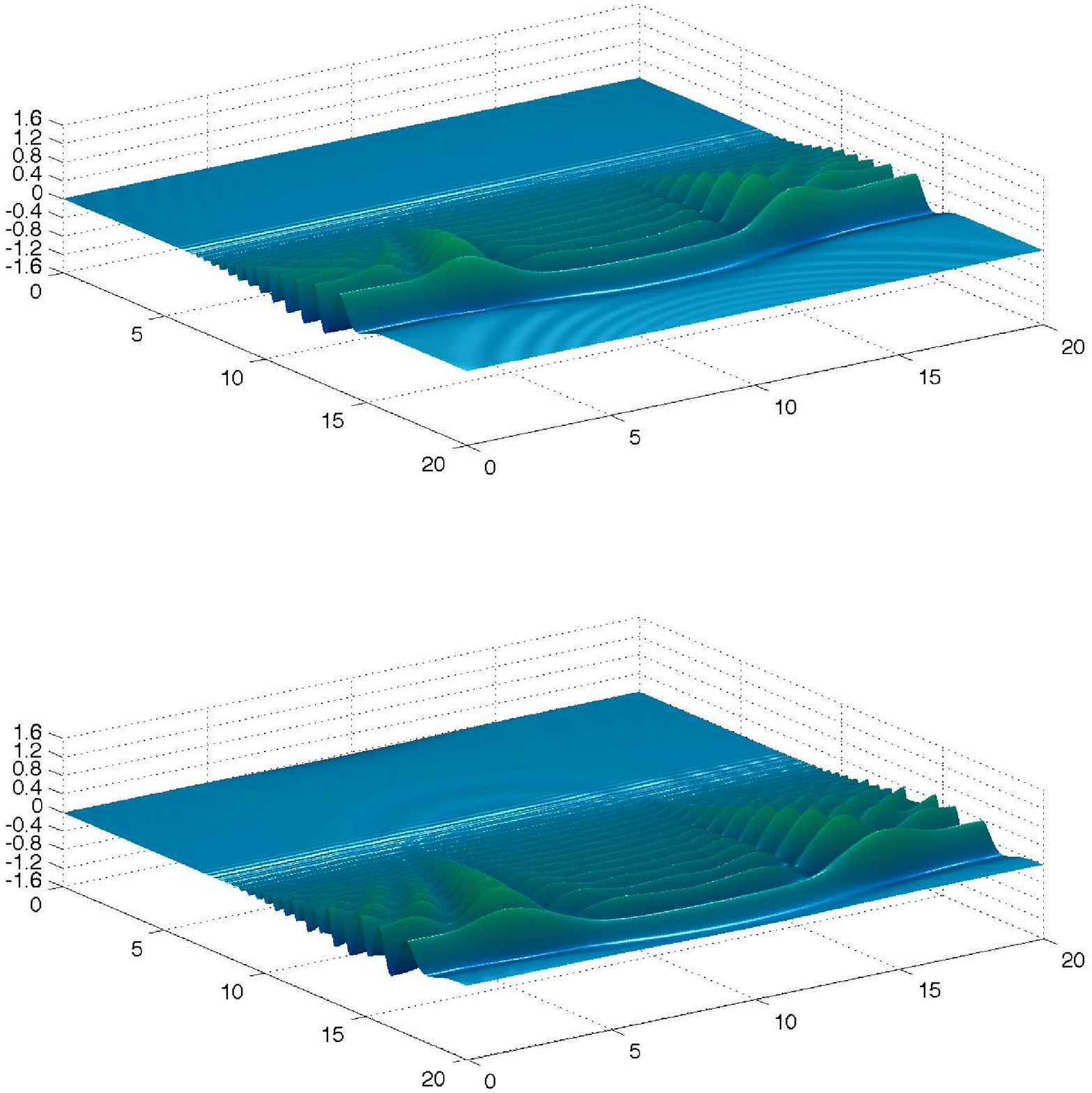}}
\caption{
Snapshots  at times $t = 13.6$ and $t=17$.  The surface elevation is propagating over a concave cavity 
(with the Luneburg lens profile; $\alpha = 0.8$)  in the $\mu = 0.1$ regime.}
\label{Fig08}
\end{figure}

In particular if the submerged Luneburg mound becomes concave, namely a circular cavity, the wavefront bending is in the other 
direction. The water becomes deeper and the wave propagates faster over the cavity.
The submerged cavity is given by $-H$ as in (\ref{S5:Eq01}). Snapshots of the  corresponding
numerical solution are given in figure \ref{Fig07} and figure \ref{Fig08}. 
At time $t = 3.1$ we have the usual incoming plane wave as shown at the top of figure \ref{Fig05}. 
At time $t = 8.6$ we start to observe the refraction pattern, now with the wavefront moving forward at its central part. 

Both simulations show that our numerical method, based on the construction of a three dimensional Dircihlet-to-Neumann 
operator, accurately captured interesting refraction patterns due to lens-like bottom topographies. In the next subsection an
approximate ray theory, for the dispersive equations, confirms the accuracy in the Luneburg lens case. 

\subsubsection{Ray theory}

We aim to compare the refraction pattern of the numerical wavefield agains  predictions from the 
theory of refraction for monochromatic waves. 
This is a classical subject in the geometrical optics theory for water waves.  A detailed treatment can be found in \citet{Whitham,Dingemans,Johnson}. Our model is reduced to three equations 
that provide a first order approximation for the refraction pattern of a nearly monochromatic wavetrain.
The monochromatic approximation is suitable for our purposes since we have a linear wave model. 
We consider the long wave regime, where dispersion is weak ($\mu$ is small).

The approximation begins with the following ansatz for the velocity potential: 
\begin{equation}\label{S5:Eq02}
\phi(x,y,t) = a(x,y)e^{i\left(\theta(x,y)-\omega t\right)}.
\end{equation}
The velocity potential is taken as an oscillatory wavetrain that is modulated with an amplitude function $a(x,y)$ and 
a phase function $\theta(x,y)$ .
Both functions $a$ and $\theta$ are assumed to vary on a scale faster than the depth  variations 
of the Luneburg lens.
The phase $\theta$  is determined  by the eikonal equation
\begin{eqnarray}
&&\theta_x^2 + \theta_y^2 = \sigma^2,\label{S5:Eq04}\\
&\mbox{where}&\omega^2 = \frac{\sigma}{\mu}\tanh(\mu(1+H(\x))\sigma),\label{S5:Eq03}
\end{eqnarray}
We have an $\bold{x}$-dependent dispersion relation that determines the variable wavenumber $\sigma(\bold{x})$. 
It plays the role of a variable index of refraction in the eikonal equation for $\theta$. 
We use the method of characteristics to solve equation (\ref{S5:Eq04}). In this context it is customary to call the 
characteristic curves $(x,y)$ as rays. It is also known that the wavefronts (level curves in $\theta$) are orthogonal 
to the rays/characteristics (\cite{Zauderer}).

The Cauchy problem for the eikonal equation consists in determining the phase function $\theta$ from an ``initial" constant data $\theta_0$ along the $y$ axis. 
First we need to determine the function $\sigma$ from the dispersion relation (\ref{S5:Eq03}). Notice that outside the circle $r=r_0$, $H$ vanishes and therefore, in that region, the solution to (\ref{S5:Eq03}) is given by a constant $\sigma_0$. Inside the circle we solve the dispersion equation numerically by means of the Newton's method.

\begin{figure}
\includegraphics[width = 0.8\textwidth]{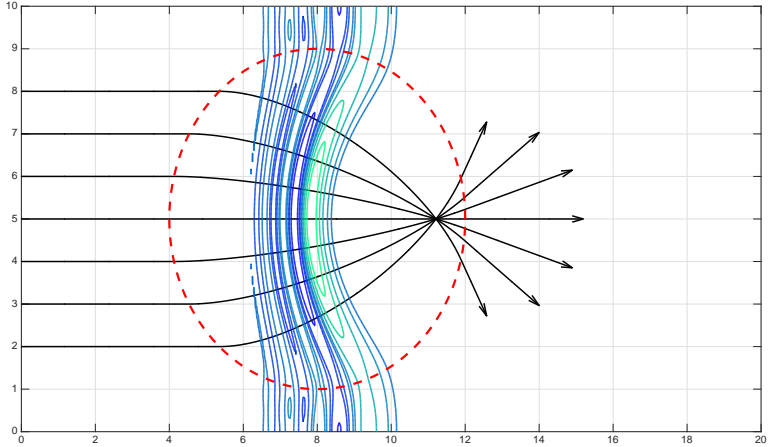}
\caption{Evolution of a Gaussian velocity potential over a Luneburg lens with parameters $\alpha = 0.8$ at time $8.6$. The black lines correspond to the rays. Notice that the wavefronts, the level lines of the velocity potential, follow the rays.
The dispersion parameter is $\mu = 0.1$}
\label{Fig09}
\end{figure}

\begin{figure}
\includegraphics[width = 0.8\textwidth]{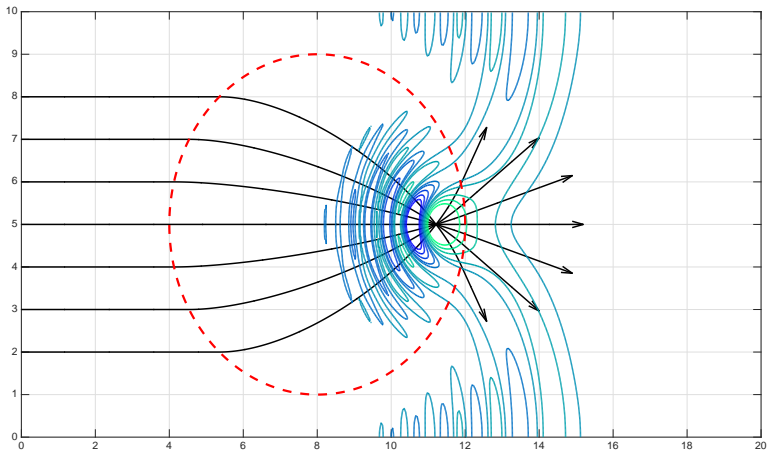}
\caption{Evolution of a Gaussian velocity potential over a Luneburg lens with parameters $\alpha = 0.8$, in the $\mu = 0.1$ regime, at time $13.6$ . The black lines correspond to the rays. Notice that the wave field focuses on top of the point where the rays meet.}
\label{Fig10}
\end{figure}

Once the function $\sigma$ and its derivatives are determined, 
we are ready to solve the characteristic equations given by:
\begin{equation}\label{S5:Eq05}
\begin{aligned}
\frac{dx}{d\tau} &=2p,\ \ \ \text{with $x(0,s)=0$.}\\
\frac{dy}{d\tau} &=2q,\ \ \ \text{with $y(0,s)=s$.}\\
\frac{dp}{d\tau} &= \frac{\partial\sigma^2}{\partial x},\ \ \ \text{with $p(0,s)=p_0(s)$.}\\
\frac{dq}{d\tau} &= \frac{\partial\sigma^2}{\partial y},\ \ \ \text{with $q(0,s)=q_0(s)$.}\\
\frac{d\theta}{d\tau} &= 2\sigma^2,\ \ \ \text{with $z(0,s)=\theta_0$.}
\end{aligned}
\end{equation}
In these equations we define  $(p,q) = \nabla \theta$ along the rays, which are parametrized by $\tau$. 
The complete set of  initial data must satisfy the following conditions, known as the strip conditions (\cite{Zauderer}):
\begin{equation}\label{S5:Eq06}
\begin{aligned}
&  p_0^2(s) = \sigma_0^2,\\
& q_0(s) = 0.
\end{aligned}
\end{equation}
The characteristic system (\ref{S5:Eq05}) is solved using the fourth-order Runge-Kutta method.
In figure \ref{Fig09} and figure \ref{Fig10} the black solid lines correspond to the rays $(x(\tau,s),y(\tau,s))$. 
Notice how the rays bend and focus into one point inside the lens, represented by the red circle. 
This focusing property distinguishes the Luneburg lens from other types of submerged mounds.
In figure \ref{Fig09} we also plot  some level curves of our numerical wave  solution, 
computed with the non-local DtN method. Notice that before the caustics, where the method of characteristics breaks down, 
the rays and wave fronts are nearly orthogonal, as expected.
Later in time, in figure \ref{Fig10}, we see the focusing of the wave field at exactly the point where the rays meet. 
The point where the rays focus through the method of characteristics is $(11.21,5)$, whereas 
the wave field, from our non-local method, attains its maximum at $(11.28,5)$. 

Our three dimensional simulations captured very well theoretical predictions. 
In the examples considered the topography had a non-trivial amplitude and our method captured remarkably well the 
predicted dynamical features.


\section{Conclusions}

We have constructed a  Dirichlet-to-Neumann (DtN) operator  for a three dimensional (surface) water wave problem in the 
presence of  highly variable, non-smooth, topographies.  The DtN operator captures the vertical structure of the flow and
allows the three dimensional problem to be reduced to a two dimensional surface. The few existing articles on the DtN 
formulation in the presence of topography consider mostly small amplitude, smooth bottom variations.  Due to memory and
numerical stability issues, the three dimensional (3D) problem had not been explored numerically. By using a Galerkin projection
onto the relevant modes that indeed interact with the topography, we have been able to exhibit several accurate simulations
in the presence of non-trivial bottom profiles. 

In particular we considered a circular submerged mound that plays the role of a Luneburg lens, known in optics. The mound
bends the wave front, through refraction, focusing the energy behind the lens.  Our 3D DtN  formulation captured quite accurately 
the focusing/refraction of the incident plane wave. Using an approximate ray theory for the dispersive (long) wave formulation
we showed that the focusing point is found with excellent accuracy.

DA gratefully acknowledges support by the PNPD/CAPES program. 
The work of AN was supported in part by CNPq under (PQ-1B) 301949/2012-2 and FAPERJ Cientistas do Nosso Estado
project no. E-26/201.164/2014. 
The authors acknowledge fruitful discussions with Paul Milewski.

\bibliographystyle{abbrv}

\end{document}